%% file: main.tex
\documentclass[prd,twocolumn,twoside,preprintnumbers,superscriptaddress,nofootinbib]{revtex4}

\usepackage{amsmath,slashed}
\usepackage{graphicx,graphics,color}
\usepackage{dcolumn}
\usepackage[hyperfootnotes=false]{hyperref}
\usepackage{xspace}
\usepackage{enumerate}
\usepackage{varwidth}
\usepackage{caption}
\usepackage{subcaption}
\usepackage{comment}

\usepackage{amssymb}
\usepackage{mathtools}
\usepackage{dsfont}
\usepackage{multirow}

\usepackage{adjustbox}
\usepackage{braket}
\usepackage{placeins}
\usepackage[T1]{fontenc}
\usepackage{float}
\usepackage{bm}
\usepackage{comment}
\usepackage{booktabs}
\usepackage{threeparttable}
\usepackage[dvipsnames]{xcolor}
\usepackage{booktabs}

\newcommand{\mN}{m_N}

\newcommand{\qq}{\mathbf{q}}
\newcommand{\pp}{\mathbf{p}}

\newcommand{\qg}{\mathbf{q}_\gamma}

\newcommand{\beq}{\begin{equation}}
\newcommand{\eeq}{\end{equation}}

\newcommand{\bea}{\begin{eqnarray}}
\newcommand{\eea}{\end{eqnarray}}

\newcommand{\bsigma}{\boldsymbol{\sigma}}

\allowdisplaybreaks[1]

\begin{document}

\title{Nucleon Electric Dipole Moments in Paramagnetic Molecules through Effective Field Theory}

\author{Wouter Dekens}
\affiliation{Institute for Nuclear Theory, University of Washington, Seattle, Washington 91195-1550, USA}
\author{Jordy de Vries}
\affiliation{Institute for Theoretical Physics Amsterdam and Delta Institute for Theoretical
Physics,University of Amsterdam, Science Park 904, 1098 XH Amsterdam, The Netherlands}
\affiliation{Nikhef, Theory Group, Science Park 105, 1098 XG, Amsterdam, The Netherlands}
\author{Lemonia Gialidi}
\affiliation{Institute for Theoretical Physics Amsterdam and Delta Institute for Theoretical
Physics,University of Amsterdam, Science Park 904, 1098 XH Amsterdam, The Netherlands}
\affiliation{Nikhef, Theory Group, Science Park 105, 1098 XG, Amsterdam, The Netherlands}
\author{Javier Men\'endez}
\affiliation{Departament de Física Quàntica i Astrofísica, Universitat de Barcelona, 08028 Barcelona, Spain}
\affiliation{Institut de Ciències del Cosmos, Universitat de Barcelona, 08028 Barcelona, Spain} 
\author{Heleen Mulder}
\affiliation{Nikhef, Theory Group, Science Park 105, 1098 XG, Amsterdam, The Netherlands}
\affiliation{Van Swinderen Institute for Particle Physics and Gravity, University of Groningen, Nijenborgh 3, 9747 AG Groningen, The Netherlands}
\author{Beatriz Romeo }
\affiliation{Department of Physics and Astronomy, University of North Carolina, North Carolina, Chapel Hill}

\begin{abstract}

Electric dipole moment (EDM) measurements using paramagnetic molecules have significantly advanced over the last decade. Traditionally, these experiments have been analyzed in terms of the electron EDM. However, paramagnetic molecules are also sensitive to hadronic sources of charge-parity (CP) violation, highlighting the need for a new framework to interpret the experimental results. 
In this Letter, we introduce an effective field theory framework to relate molecular EDMs to the EDMs of neutrons and protons. We identify the dominant contributions through power counting and pinpoint the necessary nuclear matrix elements.
As a practical application, we employ the nuclear shell model to calculate these nuclear matrix elements for the polar molecule BaF. Finally, we estimate the limits on the nucleon EDMs set by current molecular EDM experiments.
\end{abstract}

\maketitle

\section{Introduction}

Electric dipole moment (EDM) experiments are extremely sensitive probes of new sources of charge-parity (CP) violation and indirectly probe beyond-the-Standard-Model (BSM) physics at very high scales of up to $\sim 100$ TeV \cite{PospelovRitz2005,Engel:2013lsa}. Recent years have seen impressive experimental progress using polar molecules which benefit from large internal electric fields that amplify the CP-violating signal \cite{Hudson:2011zz,ACME:2013pal,Cairncross:2017fip,ACME:2018yjb,Roussy:2022cmp}. Among them, EDMs of paramagnetic systems, which have one unpaired electron, are mainly interpreted in terms of the electron EDM. Current measurements lead to a strong bound on the electron EDM, $|d_e|<4.1\cdot 10^{-30}\, e$  cm \cite{Roussy:2022cmp}, and future experiments aim to improve this by one to two orders of magnitude \cite{ACME:2018yjb,Roussy:2022cmp,NL-eEDM:2018lno,Vutha:2018tsz,Ho:2023xuo,Alarcon:2022ero,Arrowsmith-Kron:2023hcr,EuropwanEDMprojects:2025okn}. This constraint is four orders of magnitude more stringent than the neutron EDM limit \cite{Abel:2020pzs}. 

Traditionally, paramagnetic systems have not been used to constrain hadronic sources of CP violation, such as the quantum chromodynamics (QCD) $\bar \theta$ term within the SM or higher-dimensional quark-gluon operators that arise from heavy BSM physics. This is because of the assumption 
that far stricter limits can be obtained through the EDMs of the neutron or diamagnetic atoms. 
That being said, paramagnetic systems are sensitive to hadronic sources of CP violation through the CP-odd electron-nuclear force they induce \cite{Flambaum:2019ejc, Flambaum:2020gou, Mulder:2025esr}. While this force is typically strongly suppressed, the rapid progress in paramagnetic EDM experiments might make it the best way to search for hadronic sources of CP violation in the future. However, the current theoretical description of the CP-odd electron-nuclear force is still at a very rudimentary stage. 

In this Letter, we systematically derive this force as induced by the EDMs of neutrons and protons, making it possible to constrain these EDMs with paramagnetic molecular EDM experiments. As the problem involves a multitude of well-separated energy  scales, it can be systematically described using effective-field-theory (EFT) techniques. We show that this connection requires the calculation of a set of nuclear matrix elements (NMEs) that are different from the ones involved in the Schiff moments of diamagnetic systems \cite{Schiff:1963zz, Flambaum:1984fb,deJesus:2005nb}. 
As an explicit example, we compute the NMEs for the polar molecule BaF, which is being targeted by the NL-eEDM collaboration \cite{NL-eEDM:2018lno}.

\section{Effective field theory}
The calculation of molecular EDMs in terms of fundamental sources of CP violation involves widely separated energy scales. These range from the BSM and electroweak scales $(\Lambda$ and $M_W)$ to low-energy scales such as the electron mass or electron binding energy $\mathcal O(\alpha_{\mathrm{em}}^2m_e)$.  
The atomic nucleus gives rise to additional scales associated with the chiral-symmetry-breaking scale $\Lambda_\chi \sim m_N\sim 1$ GeV (comparable to the nucleon mass), the pion mass $m_\pi \sim \gamma \sim 100$ MeV (comparable to the nuclear binding momentum) and the scale of nuclear excitations $m_\pi^2/m_N \sim \mathcal O(\mathrm{MeV})$. 

Within the SM, the most relevant source of CP violation is the QCD $\bar \theta$ term, as (paramagnetic) EDMs induced by the phase in the Cabibbo-Kobayashi-Maskawa (CKM) matrix are orders of magnitude too small to be detected by current and envisioned experiments \cite{Ema:2022yra}. BSM sources are described by dimension-six operators \cite{deVries:2012ab,Dekens:2013zca,Kley:2021yhn,Kumar:2024yuu}, matched at $\Lambda_\chi$ to a $\chi$EFT Lagrangian for light mesons, nucleons, photons, electrons \cite{PospelovRitz2005,deVries:2012ab}, with the most relevant hadronic operators being quark (chromo-)EDMs, the Weinberg operator \cite{Weinberg:1989dx}, and CP-odd four-quark interactions \cite{deVries:2012ab}.
For our purposes, the most relevant interactions are given by
\begin{eqnarray}
   \mathcal{L}_{\chi} &=& \bar{g}_0 \bar{N} \tau^a N \pi^a + \bar{g}_1 \bar{N}\!N \pi^0 + \bar{g}_{0\eta} \bar{N}\!N \eta\nonumber\\
    &&+ 2 \bar N ( d_0 + d_1 \tau^3) v^\mu S^\nu N F_{\mu\nu}\,,
    \label{eq:gbarterms}
\end{eqnarray}
where the first line describes three CP-odd meson-nucleon interactions, and the second line the isoscalar and isovector nucleon EDM. 
We use the nonrelativistic nucleon doublet $N=(p,\,n)^T$ with spin $S^\mu=(0,\bsigma/2)$ and velocity $v^\mu=(1,\, \mathbf{0})$, as well as the pion triplet $\pi^a$ and the eta meson $\eta$. 

\begin{figure*}[t]
    \centering
    \begin{subfigure}[b]{0.2\textwidth}
         \centering
         \includegraphics[width=\textwidth]{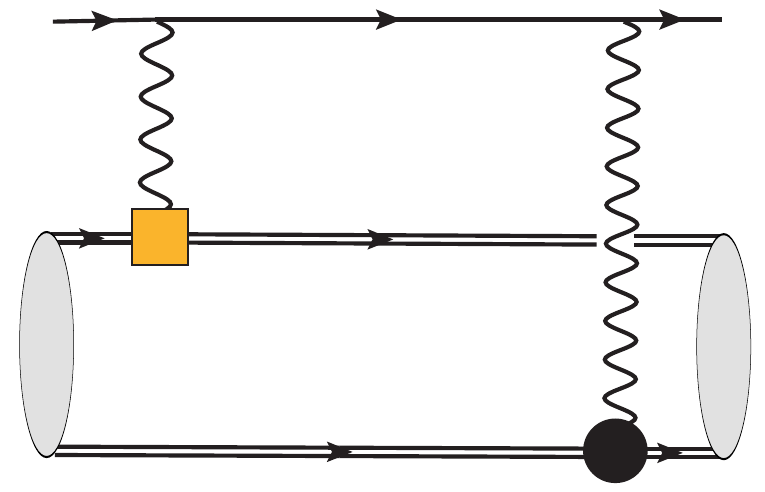}
         \caption{}
         \label{fig:2Npot}
    \end{subfigure}
    \hspace{0.05\textwidth}
    \begin{subfigure}[b]{0.183\textwidth}
         \centering
         \includegraphics[width=\textwidth]{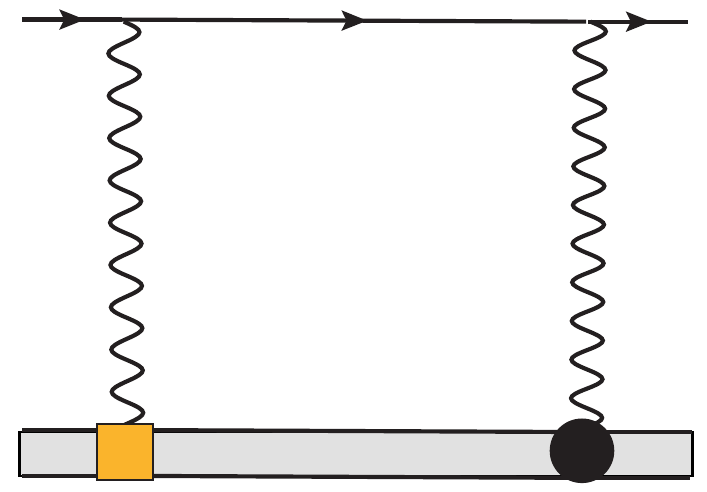}
         \caption{}
         \label{fig:usoftbox}
    \end{subfigure}
    \hspace{0.05\textwidth}
    \begin{subfigure}[b]{0.38\textwidth}
         \centering
         \includegraphics[width=\textwidth]{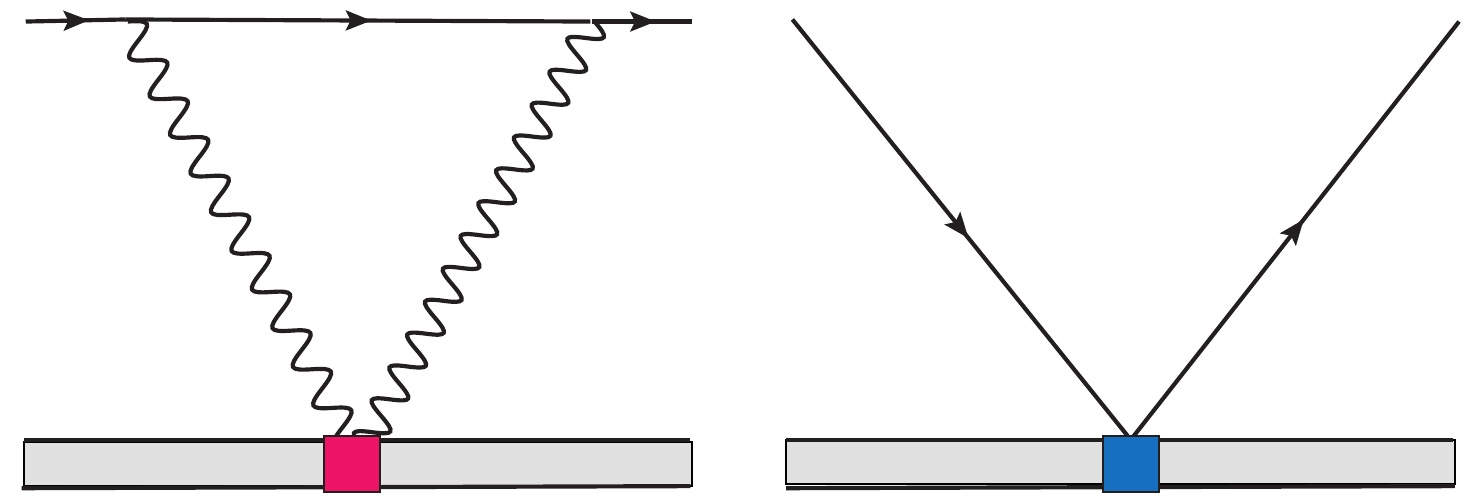}
         \caption{}
         \label{fig:betav_and_Csp} 
    \end{subfigure}
    \caption{Contributions to $\bar C_\text{SP}$ arising from the nucleon EDMs. We denote electrons by single and nucleons by double straight lines, nuclei by gray ovals (in Fig.~\ref{fig:2Npot}) or bars (in Figs.~\ref{fig:usoftbox}-\ref{fig:betav_and_Csp}), and photons by wavy lines. The black circle stands for the nucleon MDM, while the yellow, magenta, and blue squares indicate the CP-violating vertices: nucleon EDM, $\beta_v$ and $\bar{C}_\text{SP}$ effective vertices (see Eq.~\eqref{eq:CSPandbetav}), respectively. Figure \ref{fig:2Npot} shows the two-nucleon potential-region contribution, and Fig.~\ref{fig:usoftbox} the 
    ultrasoft one. Figure \ref{fig:betav_and_Csp} shows the two diagrams relevant to the matching and running of diagram \ref{fig:usoftbox} in an EFT with the nuclear ground state as the remaining degree of freedom.}
    \label{fig:boxes}
\end{figure*}

The paramagnetic EDMs induced by the meson-nucleon interactions in Eq.~\eqref{eq:gbarterms} arise mainly through intermediate CP-odd electron-nucleon interactions, which take on the form
\begin{equation}
    \mathcal{L} =  \frac{G_F}{\sqrt{2}}\bar{e}i\gamma_5 e\, \bar N\left(
    C_{\mathrm{SP}}^0 + C_{\mathrm{SP}}^1 \tau^3\right)N\,.
    \label{eq:CSPLagrangian}
\end{equation}
The nucleon EDMs in Eq.~\eqref{eq:gbarterms} give rise to contributions at longer distance scales through the diagrams in Fig.\ \ref{fig:boxes}. They induce effective interactions between the nucleus and the electrons, {\it i.e.}\ the nuclear equivalent of $C_{\rm SP}^{0,1}$, which we  denote by $\bar C_\text{SP}$, see Eq.\ \eqref{eq:CSPandbetav}.
To systematically compute the various contributions, it is useful to consider different photon modes depending on the scaling of their momentum $q_\gamma^\mu = (q_\gamma^0,\,\qg)$. We identify three regions that give relevant contributions,
\begin{enumerate}
    \item soft photons: $q^0_\gamma \sim |\qg| \sim m_\pi$,
    \item ultrasoft photons: $q^0_\gamma \sim |\qg| \sim m_\pi^2/m_N$,
    \item\label{potential} potential photons: $q^0_\gamma \sim \qg^2/\mN$, 
     $ |\qg| \sim m_\pi$,
\end{enumerate}
 and we define $Q\sim m_\pi \sim \gamma$ and $q\sim  Q^2/m_N$.

The CP-odd meson-nucleon interactions in Eq.~\eqref{eq:gbarterms} contribute to $\bar C_\text{SP}^{0,1}$ through diagrams involving a meson exchange or a pion loop in combinations with the exchange of two photons in the ultrasoft or soft region. These diagrams were first considered in Ref.~\cite{Flambaum:2019ejc} and later computed with heavy-baryon chiral perturbation theory in Ref.~\cite{Mulder:2025esr}. In addition, integrating out the mesons leads to renormalization of nucleon EDMs \cite{Crewther:1979pi,Ottnad:2009jw, Mereghetti:2010kp}, effectively shifting $d_{0,1}\rightarrow \bar d_{0,1}$, where the bar denotes the renormalized low energy constants (LECs). In what follows, we use $\bar d_{0,1}$ as the physical nucleon EDMs.

In this Letter, we focus on additional contributions to $\bar C_{\mathrm{SP}}$ from the nucleon EDMs, which arise through the topologies shown in Figs.~\ref{fig:2Npot} and \ref{fig:usoftbox}. These diagrams are captured by an effective action of the form \bea\label{eq:Seff}
&&\langle h_f (p_f)e(p_e') | iS_{\rm eff} | h_i(p_i) e(p_e)\rangle  = \frac{e^3}{2} \int_{x_i}  \langle h_f e|\\
&&\times T\Big[\bar e \slashed {A} e(x_1)\, \bar e \slashed {A} e(x_2)
\mathcal L^{(\bar d_{0,1})}_{\chi}(x_3) \, \left( A_\mu J_{\rm em}^\mu \right)(x_4) \Big]|e h_i\rangle \,,
\nonumber \eea
where we integrate over all $x_{1,2,3,4}$,  $h_{i,f}$ denote the initial and final nuclear states (for EDMs we have the nuclear ground state $|h_i\rangle = |h_f\rangle = |0^+\rangle$)  and $J_{\rm em}^\mu$ denotes the nuclear electromagnetic current. Diagrams involving nucleon EDMs and photons with soft momenta are subleading as they require the inclusion of additional pions. Power counting gives the expected size of the potential and ultrasoft  contributions
\bea
\bigg\{C_\text{SP}^\text{(pot)},\,C_\text{SP}^\text{(usoft)}\bigg\} = \frac{m_e \alpha^2 \mu_i \bar d_i}{e G_F m_N}\,\bigg\{\frac{4\pi}{Q},\,\frac{1}{q}\bigg\}\,,
\eea
where $\mu_i$ are the nucleon magnetic dipole moments (MDMs) in units of the nuclear magneton. Numerically $4\pi q \sim Q$ and these estimates are rather close, but, as we will see, they do not capture possible coherent enhancements.

\textit{Potential region:}
To evaluate the potential contributions, we can use the so-called method of regions to expand the amplitude in small ratios of scales, such as $q_\gamma^0/|\qg|$. After doing so, there are no contributions from diagrams where the nucleon EDM and $J_{\rm em}$ attach to the same nucleon (the potential region arises from picking up the poles of nucleon propagators, which can always be avoided in these one-body diagrams). There are, however, two-nucleon effects through the diagram in Fig.~\ref{fig:2Npot}. Because of spin and parity constraints, the first contributions arise from the nucleon magnetic moments, which appear in $J_{\rm em}^\mu$ at next-to-leading order ${\cal O}(Q/m_N)$. 
This results in the following contribution to the amplitude\footnote{We define $\sqrt{2E_f2 E_i}\langle h_f e | S_{\rm eff} | h_i e\rangle =(2\pi)^4\delta^4(p_f+p_e'-p_i-p_e) {\cal A}$, with the nuclear states satisfying $\langle \pp| \qq\rangle  = (2\pi)^3 \delta^3(\pp-\qq)$.}
\begin{eqnarray}
{\cal A}_{\rm pot}&=&- \langle h_f  | V | h_i \rangle \,\bar u ({p_e'} )\left(1-\frac{v\cdot (p_e'-p_e)}{2m_e}\slashed{v} \right)i\gamma_5 u({p_e})\,,
\nonumber
\eea
where we take the limit $p_e'-p_e\ll m_e$ in what follows, while $V$ denotes the potential between the two interacting nucleons. In momentum space\footnote{
Strictly speaking, the momentum space potential is infrared divergent. However, performing the Fourier transform in dimensional regularization leads to a potential in coordinate space that is IR finite. It is also possible to deal with the potential in momentum space by defining a subtraction procedure, see Appendix C of Ref.\ \cite{Cirigliano:2024msg} where a similar potential was encountered.
}
\bea\label{eq:Vq}
V &=&\frac{4e^4 m_e}{9m_N } \sum_{i\neq j}\frac{ \mu^{(i)} D^{(j)} }{|\qq|^4}\left[ \bsigma^{(i)}\cdot  \bsigma^{(j)}-\frac{1}{4}S^{(ij)}\right]
\,,
\end{eqnarray}
where $i,j$ label the nucleons, $\qq_i = \pp_i'-\pp_i$ is the exchanged momentum and we define the combination $\qq = (\qq_i - \qq_j)/2$. In addition,  $D = ( \bar{d}_0 + \bar{d}_1\tau_3)/e$, $\mu \equiv \frac{1+\kappa_0}{2}+\frac{1+\kappa_1}{2}\tau_3$ describe the EDM and MDM operators (with $\kappa_0=-0.12$ and $\kappa_1=3.7$), while $S^{(ij)} =\bsigma^{(i)}\cdot \bsigma^{(j)} -3 \qq\cdot \bsigma^{(i)}\, \qq \cdot \bsigma^{(j)}/\qq^2$. These contributions are thus determined by the NME of $V$, which scales as $1/|\qq|^4$. The form of this two-body potential is similar to the
NMEs appearing in neutrinoless double-$\beta$ decay \cite{Agostini:2022zub} or radiative corrections to superallowed $\beta$ decays \cite{Cirigliano:2024msg,Cirigliano:2024rfk} although with different isospin and/or $\qq$ dependence. The many-body techniques needed to compute such NMEs, including {\it ab initio} approaches, have developed significantly in the last decade and can be directly applied to these EDM calculations \cite{Agostini:2022zub,Cirigliano:2022rmf}.

\textit{Ultrasoft region:}
In this region, we expand in small ratios of scales, such as  $q_\gamma^0/m_\pi\sim |\qg|/m_\pi$ and $ m_e/m_\pi$. Photons with this (small) momentum scaling can be thought of as  coupling to the nucleus as a whole, instead of the individual nucleons. After inserting a complete set of (nuclear) states between the hadronic operators and working out the time-ordered product in Eq.\ \eqref{eq:Seff}, we again find that the leading contributions involve the magnetic moments,
\begin{widetext}
\begin{align}\label{eq:usoftMag}
{\cal A}_{\rm usoft}&= \frac{-2i e^4}{m_N}  \int \frac{d^4k}{(2\pi)^4} \frac{\bar u(p_e')\gamma_\lambda (\slashed {k}+m_e)\gamma_\rho u(p_e)\,\epsilon_{\sigma \alpha \beta\eta}v^\beta}{(k^2-m_e^2)(k-p_e)^2(k-p_e')^2} \sum_n\Bigg[(p_e-k)^\sigma (p_e'-k)^\mu g^{\nu \lambda}g^{\alpha\rho}\nonumber\\
&
\times \left(\frac{ \langle h_f| D_{\mu\nu}|n\rangle\langle n|M^\eta |h_i\rangle}{v\cdot l_+-E_n +i\epsilon}
+\frac{ \langle h_f| M^\eta |n\rangle\langle n |D_{\mu\nu}|h_i\rangle}{v\cdot l_--E_n +i\epsilon}
\right)+ \Big\{p_e\leftrightarrow p_e',\,l_+\leftrightarrow l_-,\,\alpha\leftrightarrow \nu \Big\}\Bigg]\,,
\end{align}
\end{widetext}
where $|n\rangle$ denote intermediate nuclear $|1^+\rangle$ states with energies $E_n$, while $l_{+} = p_i+p_e-k$,  $l_{-} = p_i-p_e'+k$. Furthermore,  $M^\eta\equiv \bar N \mu \, S^\eta N(0)$ and
$D_{\mu\nu} = \bar N D \left(v_\mu S_\nu-v_\nu S_\mu\right)N(0)$ denote the MDM and EDM operators. 

Although the appearing integrals can  be evaluated using known techniques \cite{Zupan:2002je}, the expressions are rather unwieldy. They greatly simplify if there is a hierarchy between the nuclear excited states and the electron momenta, $p_e\sim m_e\ll \Delta_n=|E_n-E_i|$, which is a good approximation for $^{138}$Ba as discussed below. Likewise, the relevant excited states in magnetic-dipole transitions --- also driven by the spin operator --- in isotopes of Yb, Hf, and Th with an even number of neutrons also enter at about 2~MeV or higher energies~\cite{Heil:1988esp,Pietralla:1997bfy,Zilges:1990ezd}. In this case, Eq.\ \eqref{eq:usoftMag} can be captured by  a low-energy nuclear EFT in which the excited nuclear states have been integrated out, but which still contains electrons, ultrasoft photons, and the ground state of the nucleus. The relevant interactions in this theory can be written as 
\bea\label{eq:CSPandbetav}
{\cal L}_{\Psi} =\Psi_i^\dagger\left[ \frac{G_F}{\sqrt{2}}\bar C_\text{SP}  \, \bar e i\gamma_5 e +\beta_v\,v^\alpha F_{\alpha \beta} v_\lambda  \epsilon^{\beta\lambda\mu\nu}F_{\mu\nu}\right] \Psi_i\,,
\eea
where $\Psi_i$ denotes the spin-0 field describing the nucleus\footnote{We describe the nucleus nonrelativistically, so that the kinetic term takes the form ${\cal L}_{\Psi}^{(0)} = \Psi_i^\dagger i v\cdot D \Psi_i$. This ensures  that $\langle 0| \Psi_i(x) |\pp\rangle = e^{-i p\cdot x}$ and implies the field has dimension $[\Psi_i]=3/2$, so that $[\bar C_\text{SP} ]= 0$.},
$\bar C_\text{SP}$ describes the nuclear version of $C_\text{SP}^{0,1}$, while $\beta_v$  has a similar form as the nuclear polarizability but violates CP. At the scale $\lambda=\Delta_n$, $\beta_v$ obtains a contribution from integrating out the excited states at tree level, while  $\bar C_\text{SP}$ arises from Eq.\ \eqref{eq:usoftMag}. 
After expanding in   $m_e/ |E_n-E_i|$, this expression simplifies and the remaining integrals 
can be calculated using techniques from Ref.~\cite{Broadhurst:1991fz}.
All in all, matching the nucleon-level theory to the EFT without excited states then gives at a scale $\lambda\simeq \Delta_n$,
\bea
  \frac{G_F}{\sqrt{2}} \bar C_\text{SP}&=& -\frac{e^4m_e}{4\pi^2 m_N}\sum_n\frac{A_n}{\Delta_n}\left(4-3\log\frac{4\Delta _n^2}{\lambda^2}\right)\,,\\
\beta_v &= &\frac{e^2}{m_N}\sum_n \frac{A_n}{\Delta_n}\,,\quad 
A_n =- \frac{\langle h_i | D \bsigma | n \rangle \cdot \langle n |  \mu \bsigma |h_i \rangle}{12}\,.\nonumber
\eea

These interactions can be evolved from $\lambda\sim \Delta_n$ to lower energies, $\lambda_e\sim m_e$, using the renormalization group equation which arises through the loop diagram of Fig.~\ref{fig:betav_and_Csp}. 
The amplitude at low scales, $\lambda_e\sim m_e$, can finally be expressed as the sum of $\bar C_\text{SP}(m_e)$ and a loop involving $\beta_v$. 

We capture the total combination of the ultrasoft and potential contributions by an effective contact interaction, such that ${\cal A}_{\rm total} = \frac{G_F}{\sqrt{2}} \bar C_\text{SP}^{\rm eff} \bar u(p_e') i\gamma_5 u(p_e)$ with 
\begin{align}\label{CSPeffTOT}
\bar C_\text{SP}^{\rm eff} =-\frac{\sqrt{2}}{G_F}\left[\frac{4\alpha^2 m_e}{m_N}\sum_n \frac{A_n}{\Delta_n}\left(3\ln \frac{m_e^2}{4\Delta_n^2}-1\right)+ \langle h_i| V|h_i\rangle\,\right]\,,
\end{align}
which  
is independent of the renormalization scale $\lambda$. Additional details of the derivation can be found in the Supplemental Material. We stress that this is the effective interaction between electrons and the nucleus as a whole and differs from Eq.~\eqref{eq:CSPLagrangian}, which is the coupling to individual nucleons. Evaluating the ultrasoft region thus requires the excited state energies, $\Delta_n$, and the set of nuclear matrix elements of the one-body operator $\sim \langle h_i |\bsigma |n\rangle$ contained in $A_n$. These have a form similar to the leading two-neutrino double $\beta$, and double magnetic-dipole NMEs \cite{Simkovic:2018rdz,Morabit:2024sms,Romeo:2021zrn} and of subleading NMEs of the neutrinoless double-$\beta$ decay~\cite{Dekens:2024hlz,Castillo:2024jfj}. Therefore, similar many-body methods used in these studies can be applied here. 
Equation \eqref{CSPeffTOT} is the main result of this work and makes it possible to connect nucleon EDMs to measurements of paramagnetic molecules.

\section{Nuclear matrix elements} We now focus on the polar molecule BaF, which is targeted by the NL-eEDM collaboration \cite{NL-eEDM:2018lno}. The heaviest atom in the molecule, ${}^{138}$Ba, has a magic neutron number, and it is just two neutrons away from ${}^{136}$Ba, the well-studied~\cite{Caurier:2011gi} final state of the double-$\beta$ decay of ${}^{136}$Xe. 
We calculate excitation energies and NMEs with the nuclear shell model~\cite{Caurier:2004gf}, using the codes~\cite{shimizu2013,TMiyagiLib}, a configuration space of $1d_{5/2}$, $0g_{7/2}$, $2s_{1/2}$, $1d_{3/2}$, $0h_{11/2}$ orbitals with a $^{100}$Sn core, and three effective interactions: GCN5082~\cite{Caurier:2010az}, QX~\cite{QiQX} and Sn100pn~\cite{Brown05}; details are in the Supplemental Material. The value of the ultrasoft NME is 
\beq\label{usoftBa}
\bar C_\text{SP}^{\rm usoft} =  \left(67 \pm 28\right) \,  d_p\, (e\,\mathrm{fm})^{-1}\,,
\eeq
where $d_p = \bar d_0 + \bar d_1$. We only find sensitivity to the proton EDM because, in our calculation, the 82 neutrons form a closed shell, as $^{138}$Ba is magic in neutrons. This is also why the first intermediate $1^+$ excited state appears around 2.5 MeV. The largest contribution to the ultrasoft NME arises from states around $E_n = 4.5\,\mathrm{MeV} \gg m_e$, justifying our approximation, and higher-energy states only contribute mildly. We show the cumulative contribution from the excited-state spectrum in the Supplemental Material.

The potential contribution evaluates to
\beq \label{potBa}
\bar C_\text{SP}^{\rm pot} = \left[(-433 \pm 5) \, d_p + (387\pm 0.4)\, d_n \right]\,(e\, \mathrm{fm})^{-1}\,,
\eeq
where $d_{n} = \bar d_0 - \bar d_1$. 
Compared to the ultrasoft regime, the potential contribution is dominant. This is because of the coherent nature of the potential NME, which leads to a linear scaling with the total number of protons, $Z$ ($d_p$ term), or neutrons, $N$ ($d_n$ term), in the nucleus. The coherence appears as most NME contributions stem from proton-proton and neutron-neutron spin-zero pairs, prevalent in nuclei due to the attractive pairing interaction, and also dominant for the long-range potential (proportional to $|\mathbf{r_i-r_j}|$ in coordinate space). 
This scaling is in rough agreement with the estimate of Ref.\ \cite{Flambaum:2019ejc}, as well as an evaluation of the potential of Eq.\ \eqref{eq:Vq} in a Fermi gas state\footnote{We thank J. Engel for discussions on this point.}.
Our many-body calculations, which also cover nuclei lighter than $^{138}$Ba,  suggest that nucleus-dependent effects can correct this estimate by up to $20\%$. 
The coherent character makes potential NMEs less dependent on the details of the nuclear structure, reducing their relative uncertainty with respect to ultrasoft NMEs (the very small error in the $d_n$ potential NME is because in our calculation the 82 neutrons form a closed shell). We provide more details in the Supplemental Material. 
While we do not expect any breakdown of the scaling behavior discussed above for the potential contribution, in nuclei with particularly strong low-lying M1 excitations, the ultrasoft contributions could be enhanced\footnote{Such enhancements arise, for example, for sterile neutrino contributions to neutrinoless double beta decay \cite{Dekens:2024hlz}.}.
 Explicit calculations of ultrasoft and potential NMEs in heavier systems such as Th or Hf would be very interesting.

The uncertainties in Eqs.\ \eqref{usoftBa} and \eqref{potBa} were estimated by considering the differences between several nuclear Hamiltonians, see the Supplemental Material for details. 
In addition, a fully quantified theory error should take into account the errors due to the many-body method and missing orders in chiral EFT. The latter are expected to give ${\cal O}(m_\pi/\Lambda_\chi)$ corrections, while more solid estimates require explicit calculations, e.g.\ by computing the corrections to the potential in Eq.\ \eqref{eq:Vq}. We leave a more complete analysis of theory uncertainties to future work.

The expected sensitivity of the BaF experiment  is an electron EDM equivalent of $d_e \leq 10^{-30}$ e cm \cite{NL-eEDM:2018lno}. Using the NME calculations of this Letter, this would correspond to a sensitivity to the nucleon EDMs $|d_p|_{\text{BaF}}<8.4 \cdot 10^{-24}\;e\,\mathrm{cm}  $ and $|d_n|_{\text{BaF}}<8\cdot 10^{-24}\;e\,\mathrm{cm}  $.
While we do not have shell-model calculations for the most precise experiment based on HfF$^+$, we can use the linear $Z$ and $N$ dependence of the potential NME, where nucleus-dependent corrections are expected at the $\sim 20\%$ level, with additional higher-order chiral corrections of $O(m_\pi/\Lambda_\chi)$, to estimate 
\begin{equation}
    |d_p|_{\text{\tiny HfF}^+}\lesssim 1.6 \cdot 10^{-23}\;e\,\mathrm{cm}\,,\,|d_n|_{\text{\tiny HfF}^+}\lesssim 1.6 \cdot 10^{-23}\;e\,\mathrm{cm}\,,
\end{equation}
roughly two orders of magnitude weaker than the proton EDM limit set by $^{199}$Hg \cite{Graner_2016} and three orders than the direct neutron EDM limit \cite{Abel:2020pzs}. Both of these limits have been stable within an order of magnitude in the last decade. Based on the prospects for direct neutron EDM experiments \cite{Alarcon:2022ero}, an order of magnitude improvement can be expected. Experiments with radioactive molecules could provide very stringent limits, but are still under development 
\cite{Arrowsmith-Kron:2023hcr}. Considering the past and anticipated progress in paramagnetic polar molecular EDM experiments, with projected improvements of two to three orders of magnitude within a decade \cite{Alarcon:2022ero}, the precision gaps in the limits on $|d_p|$ and $|d_n|$ are not insurmountable.

\section{Discussion}
A nonzero EDM measurement raises the question of the underlying CP-violating mechanism. Ratios of paramagnetic systems disentangle electron EDM from CP-odd electron-nucleon contributions \cite{Chupp:2014gka, Fleig:2018bsf}, and nuclear-to-nucleon EDM ratios separate quark-gluon sources \cite{deVries:2011re}. We devise a complementary strategy to identify the underlying source of CP violation.
Besides the diagrams calculated in this Letter, there appear contributions from meson-exchange diagrams \cite{Flambaum:2019ejc, Mulder:2025esr}. Depending on the underlying CP-violating source, the ratio of meson to nucleon EDM contributions varies. We can be most concrete for the QCD $\bar \theta$ term, where the sizes of the low-energy constants in Eq.~\eqref{eq:gbarterms} are known relatively well \cite{Bsaisou:2012rg,deVries:2015una,Richardson:2025rnm}. For BaF, the meson diagrams give a contribution
$\bar C^{\mathrm{meson}}_\text{SP}\text{(BaF)} =  (220
\pm 62)\cdot 10^{-2}\,\bar \theta$ \cite{Mulder:2025esr}. 
Inserting lattice-QCD nucleon EDMs 
\cite{Dragos:2019oxn,Liang:2023jfj} gives \beq
\bar C^{\mathrm{pot+usoft}}_\text{SP} = -(196\pm 54)\cdot 10^{-2}\,\bar \theta\,.  
\eeq
comparable in magnitude to the meson-exchange diagrams but with \textit{opposite} sign --- an accidental cancellation specific to the $\bar \theta$ term.
We combine all contributions to compute the \textit{equivalent} electron EDM $d_e^\text{equiv} \equiv r_\text{mol} \bar C_\text{SP}/A$ \cite{Pospelov:2013sca}, which is convenient as paramagnetic EDM searches are usually interpreted as limits on $d_e$. 
For BaF, $r_\text{BaF} = 4.46[18]\cdot 10^{-21} e\text{ cm}$ \cite{Haase_2021}, and we obtain $d_e^\text{equiv}(\bar{\theta}) = (7.5 \pm 27) \cdot 10^{-24} \, \bar{\theta} \,e \text{ cm}$.

\begin{figure}[t!]
    \centering
    \includegraphics[width=\linewidth]{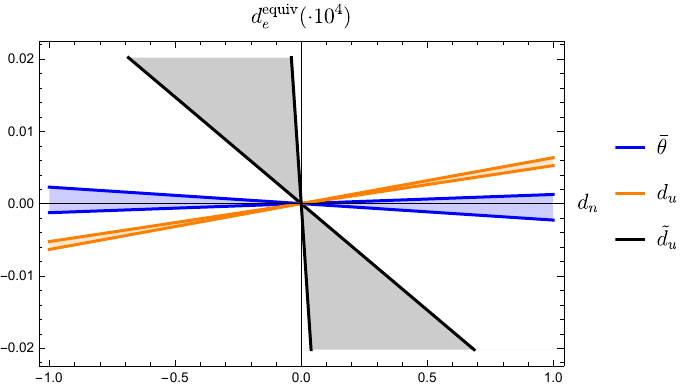}
    \caption{Ratio $d_e^\text{equiv}/d_n$ as a function of $d_n$ (in arbitrary units), for three underlying CP-violating sources: the $\bar{\theta}$ term (blue band), the up quark EDM $d_u$ (orange) or chromo-EDM $\tilde{d}_u$ (grey). Each band shows the predicted ratio assuming a given source; the width reflects hadronic uncertainties. The QCD matrix elements of Refs.~\cite{Dekens:2018bci,Pospelov:2001ys,Bhattacharya:2025blb} were used to connect the CP-violating sources to the hadronic couplings, $d_{n,p}$ and $\bar g_{0,1,0\eta}$.} 
    \label{fig:ratioplot}
\end{figure} 

For other hadronic sources of CP violation, power counting arguments give insight into the ratio of meson-to-nucleon-EDM contributions. For example, the quark chromo-EDM breaks CP and isospin symmetry. The meson-nucleon interactions in Eq.~\eqref{eq:gbarterms} are the leading CP-violating hadronic interactions \cite{Pospelov:2001ys,deVries:2012ab} and their contributions dominate the paramagnetic EDMs. On the other hand, for quark EDMs and the Weinberg operator the mesonic interactions are suppressed by, respectively, $\alpha/\pi$ (electromagnetic suppression)  and $m_\pi^2/\Lambda_\chi^2$ (chiral suppression) \cite{deVries:2012ab}. As such, the ratio of $\bar C^{\mathrm{eff}}_\text{SP}$ (and thus $d_e^\text{equiv}$) to the neutron EDM is different. We illustrate this in Fig.~\ref{fig:ratioplot}, where we plot the  $d_e^\text{equiv}$, including both mesonic and nucleon EDM contributions, against $d_n$. The bands correspond to scenarios where $d_e^\text{equiv}$ and $d_n$ are sourced, respectively, by the $\bar{\theta}$ term, the up quark EDM $d_u$, and the up quark chromo-EDM $\tilde{d}_u$. Remarkably, the ratio of neutron to \textit{paramagnetic} EDMs can identify the underlying hadronic source of CP violation. We emphasize that this identification is independent of the numerical value of the CP-odd source, be it $\bar \theta$, $d_q$ or $\tilde d_q$, as these drop out in the ratio. As an example, consider a scenario where $d_n$ is measured at $10^{-27}\, e\,$cm in a future experiment. Figure \ref{fig:ratioplot} shows that if $\tilde{d}_u$ is the underlying CP-odd source, we would expect $d_e^\text{equiv}$ to be negative and of order $10^{-32}-10^{-33}\,e\,$cm, while $d_u$ should lead to a positive $d_e^\text{equiv}$ of order $10^{-34}\,e\,$cm, and $\bar \theta$ to even smaller values.

\section{Conclusions} The EFT approach presented in this Letter allows one to derive the contribution from nucleon EDMs to paramagnetic EDMs in a systematic way. 
We have identified the novel NMEs needed to interpret paramagnetic EDMs in terms of hadronic CP violation.
While power counting arguments indicate that similar contributions 
would arise from potential and ultrasoft virtual photons, explicit shell-model calculations show that potential NMEs dominate because of the coherent contribution of most protons and neutrons in the nucleus.
Experimental improvements of two to three orders of magnitude in paramagnetic molecular systems are needed to set competitive limits on nucleon EDMs.  Finally, we have shown that ratios of paramagnetic-to-neutron EDMs 
point toward the underlying mechanism of CP violation. 

\begin{acknowledgements} 
J.d.V. and H.M. thank our colleagues from the NL-eEDM collaboration for discussions and encouragement. We thank Jon Engel, Emanuele Mereghetti, and Rob Timmermans for important discussions. This work was partly funded by the Netherlands Research Council (NWO) under Program No. XL21.074, and by 
MCIN/AEI/10.13039/501100011033 from the following grants: PID2023-147112NB-C22; CNS2022-135716 funded by the “European Union NextGenerationEU/PRTR,” and CEX2024-001451-M  to the “Unit of Excellence Mar\'ia de Maeztu 2025-2031” award to the Institute of Cosmos Sciences; and by the Generalitat de Catalunya, through Grant No. 2021SGR01095. B. R. acknowledges support from the U.S. Department of Energy under Contract No. DE-FG02-97ER4101. W.D. was supported by the INT’s U.S.
Department of Energy Grant No. DE-FG02-00ER41132.  
\end{acknowledgements} 

\appendix
\input{finalSM}

\bibliography{reference}

\end{document}

%% file: finalSM.tex
\section{Ultrasoft contribution to $\bar{C}^{\rm eff}_{SP}$}
\label{sec:radial}
The integrals relevant for the ultrasoft contributions take the form
\beq I_n \equiv \int \frac{d^dk}{(2\pi)^d} \frac{1}{(k^2)^n}\frac{1}{v\cdot k-\Delta}\,, \eeq
which can be calculated using techniques from Ref.~\cite{Broadhurst:1991fz}:

\bea
I_n = 2i \frac{(-1)^{n+1}}{(4\pi)^{d/2}}\frac{\Gamma(2n+1-d)\Gamma(\frac{d-2n}{2})}{\Gamma(n) (2\Delta)^{1+2n-d}}\,.
\eea
This can be used to evaluate the ultrasoft amplitude ${\cal A}_{\rm usoft}$, whose effects are encoded in a contribution to the CP-odd electron-nucleus interaction
\begin{equation}
    \bar{C}^{\rm usoft}_{SP}=-\frac{\sqrt{2}\alpha^2 m_e}{3  m_N G_F}M^{\rm usoft}_{SP}\;,
    \label{eq:Csp}
\end{equation}
where $M_{SP}^{\rm usoft}$ is the NME between initial and final nuclear states, $\ket{h_{i,f}}$, defined by
\begin{equation}
    M^{\rm usoft}_{SP}=\sum_n \frac{ \langle h_f |  D^{(i)}\vec \sigma | n \rangle \cdot \langle n |\mu^{(i)}\vec \sigma|h_i \rangle}{\Delta_n}\left(1+3\ln \frac{4\Delta_n^2}{m_e^2}\right)\;
    \label{eq:Msp}.
\end{equation}
Here $\Delta_n=E_n-E_i$ is the excitation energy of the intermediate nuclear states $\ket{n}$, and the nucleon EDMs and MDMs $D^{(i)}= (\bar d_0 + \bar d_1 \tau_3^{(i)})/e$ and $\mu^{(i)}=(\mu_0 +\mu_1 \tau_3^{(i)})$ are defined in terms of the isoscalar and isovector nucleon EDMs, $\bar{d}_{0,1}$, and the isoscalar and isovector anomalous magnetic moments, $\kappa_{0,1}$, through $\mu_{i}=(1+\kappa_{i})/2$. 

We focus on $^{138}$Ba, the heaviest nucleus in the diatomic polar molecule BaF used  by the NL-eEDM experiment~\cite{NL-eEDM:2018lno}. We use the nuclear shell model to compute the matrix elements involving the one-body spin operator in Eq.~\eqref{eq:Msp} to the set of $1_n^+$ nuclear excited states, as well as the relevant excited-state energies. Nonetheless, we have adjusted these energies to exactly reproduce the one of the first $1^+$ excited state of $^{138}$Ba. This changes $M^{\rm usoft}_{SP}$ just by 2\%-7\% 
depending on the effective interaction used. 

Since in our calculation for $^{138}$Ba the 82 neutrons completely fill the configuration space, we cannot create any particle-hole excitation involving a neutron orbital, meaning there is no sensitivity to $d_n$.
Therefore, for $^{138}$Ba Eq.~\eqref{eq:Msp} reduces to
\begin{align}
    M^{\rm usoft}_{SP}=&(\bar{d}_0+\bar{d}_1)(\mu_0+\mu_1) m_{\sigma}=m_{\sigma}\mu_{p}d_{p}\;,
\end{align}
with $2\mu_{p}=\kappa_0+\kappa_1+2$ (likewise, $2\mu_{n}=\kappa_0-\kappa_1$) and $m_{\sigma}$ defined by
\begin{equation}
    m_{\sigma}=-\frac{1}{e}\sum_n \frac{\bra{0^+_1}\vec{\sigma}\ket{1_n^+}^2}{\Delta_n}\left(1+3\ln \frac{4\Delta_n^2}{m_e^2}\right)\;,
\label{eq:Ms}
\end{equation}
where we drop the isospin operator as only protons contribute to the NME. As indicated by Eq.~\eqref{eq:Ms}, different contributions cannot cancel.

\begin{table}[t]
\centering
\scalebox{1.}{
\def\arraystretch{1.3}
\begin{tabular}{cc ccc}
\hline \hline
 &  & GCN5082 & QX & Sn100pn \\
\hline
\multirow{2}{*}{$^{138}$Ba} 
 &  $d_p$  & 61.0 &  97.0 & 41.7  \\
 &  $d_n$  & 0 &  0 & 0  \\
\hline
\multirow{2}{*}{$^{106}$Sn} 
 &  $d_p$  & 0 &  0 & 0  \\
 &  $d_n$  & -90.0 & -89.6 & -55.5 \\
\hline
\multirow{2}{*}{$^{104}$Te} 
 &  $d_p$  & 55.3 & 66.0 & 53.8  \\
 &  $d_n$  &  -43.7 &  -54.8 &  -46.9  \\
\hline
\multirow{2}{*}{$^{132}$Te} 
 &  $d_p$  & 11.9 & 11.0 & 6.5  \\
 &  $d_n$  &  -24.9 &  -16.2 &  -30.2  \\
\hline\hline
\end{tabular}
}
\caption{$\bar{C}^{\rm usoft}_{SP}$ results for various nuclei with similar mass number as $^{138}$Ba, obtained with three different shell-model Hamiltonians~\cite{Caurier:2010az,QiQX,Brown05}. Here units are $(e\, \mathrm{fm})^{-1}$. }
\label{tab:calc_Csp_nuclei}
\end{table}

\begin{figure}[t]
    \centering
    \includegraphics[width=0.45\textwidth]{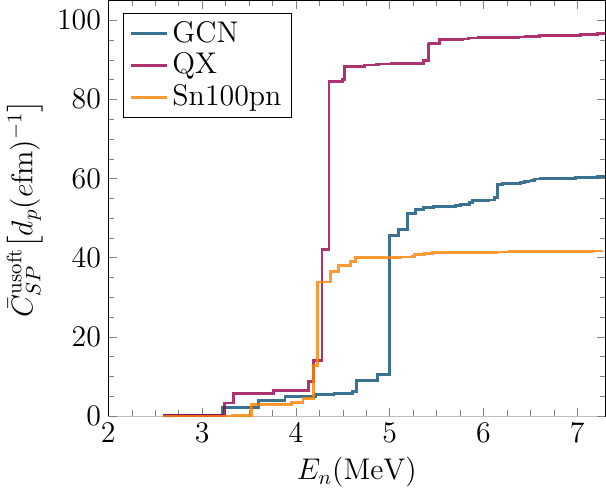} 
    \caption{$\bar{C}^{\rm usoft}_{SP}$ as a function of the excitation energy of the intermediate states, for three shell-model Hamiltonians.}
    \label{fig:Csp_running_sum}
\end{figure}

Table~\ref{tab:calc_Csp_nuclei} presents the calculated 
$\bar{C}^{\rm usoft}_{SP}$ values for $^ {138}$Ba. The results obtained with the three different shell-model Hamiltonians used differ, at most, by about a factor two. This shows a significant sensitivity to nuclear structure for the ultrasoft NME. 
We note that, similar to electromagnetic transitions, our valence-shell calculations may require an effective ultrasoft operator~\cite{Brown05}, which should be studied in future work. Nonetheless, we expect this effect to be smaller than the error due to the nuclear Hamiltonian given in Table~\ref{tab:calc_Csp_nuclei}.

Figure~\ref{fig:Csp_running_sum} shows the cumulative sum of $\bar{C}^{\rm usoft}_{SP}$ 
as a function of the excitation energy of the intermediate states. 
For the three nuclear Hamiltonians used, the behaviour is quite similar: a few states between $4-5$~MeV dominate, with lower- and higher-energy states contributing little.

Additionally, Table~\ref{tab:calc_Csp_nuclei} also presents the results for the ultrasoft NME in other nuclei, $^{106}$Sn, $^{104}$Te and $^{132}$Te, using the same configuration space as for $^{138}$Ba. Our results indicate NME values comparable to the $^{138}$Ba ones. The theoretical uncertainty due to the nuclear Hamiltonian used is also comparable to the one found for $^{138}$Ba, highlighting again the sensitivity of the NME to nuclear structure effects.

\section{Potential contribution to $\bar{C}^{\rm eff}_{SP}$}
\label{sec:nmes}
In the potential region the contribution to the CP-violating electron-nucleus interaction is given by 
\begin{equation}
    \bar{C}^{\rm pot}_{SP}=-\frac{\sqrt{2}}{G_F}\langle h_i | V | h_i \rangle\;,
\label{eq:Csp_pot_ori}
\end{equation}
where in coordinate space 
\begin{equation}
    V(\vec{r})=-\frac{e^4m_e}{18\pi m_N}\sum_{i\neq j}\mu^{(i)}D^{(j)}\,|\vec{r}|\left[\sigma^{(i)}\cdot\sigma^{(j)}+\frac{1}{16}S^{(ij)}(r) \right],
\end{equation}
with $S^{(ij)}=3\hat{r}\cdot\vec{\sigma_i}\hat{r}\cdot\vec{\sigma_j}-\vec{\sigma_i}\cdot\vec{\sigma_j}$. For convenience we define $N^{(ij)}=\sigma^i\cdot\sigma^j+S^{(ij)}/16$ and rewrite Eq.~\eqref{eq:Csp_pot_ori} as
\begin{equation}
    \bar{C}^{\rm pot}_{SP}=\frac{\sqrt{2}e^4m_e}{18\pi G_Fm_N}M^{\rm pot}_{SP},
    \label{eq:Csp_pot}
\end{equation}
with $M^{\rm pot}_{SP}$ the expectation value of the operator 
\begin{align}
   O_{SP}&=\;\frac{\bar{d}_0}{e}\sum_{i\neq j}r\left(\mu_0+\frac{\mu_1}{2}(\tau_3^i+\tau_3^j)\right)N^{(ij)}\nonumber\\
   &+\frac{\bar{d}_1}{e}\sum_{i\neq j}r\left(\mu_1\tau_3^i\tau_3^j+\frac{\mu_0}{2}(\tau_3^i+\tau_3^j)\right)N^{(ij)}\;,\label{eq:Osp_pot_d0_d1}
\end{align}
or in terms of $d_{p,n}$,
\begin{align}
   O_{SP}&=\frac{d_p}{2e}\sum_{i\neq j}r\left(\mu_0+\mu_1\tau_3^i\tau_3^j+\frac{\mu_1+\mu_0}{2}(\tau_3^i+\tau_3^j)\right)N^{(ij)}\nonumber\\
   &+\frac{d_n}{2e}\sum_{i\neq j}r\left(\mu_0-\mu_1\tau_3^i\tau_3^j+\frac{\mu_1-\mu_0}{2}(\tau_3^i+\tau_3^j)\right)N^{(ij)}\label{eq:Osp_pot_dp_dn}.
\end{align}

For calculating the two body matrix elements of this operator, $M_{SP}^{\rm pot}=\bra{0^+}O_{SP}\ket{0^+}$, we have modified the codes NATHAN~\cite{MPinedo} and \texttt{imsrg++}~\cite{Ragnar_2018}, where similar operators are calculated for double-$\beta$ decays. 
We note that both core and valence nucleons contribute to the potential NME. In fact, in the ideal case of a nucleus with fully-closed angular-momentum shells for both neutrons and protons, the $\sum_{i\ne j}\sigma^{(i)}\cdot\sigma^{(j)}$ operator with the same isospin dependence as in Eq.~\eqref{eq:Osp_pot_d0_d1} would just count the number of proton pairs and neutron pairs coupled to spin zero. The corresponding NME is $(-3 Z\,\mu_p d_p-3 N\,\mu_n d_n) (\mathrm{fm}/e)$, with separate coherent contributions of all protons and all neutrons in the nucleus. 

\begin{table}[t]
\scalebox{1.0}{
\centering
\begin{tabular}{c cc cc cc}
\hline\hline
\addlinespace[0.8ex]
    & \multicolumn{2}{c}{$\sigma^i\sigma^j$} 
    & \multicolumn{2}{c}{$S^{ij}$} 
    & \multicolumn{2}{c}{$\sigma^i\sigma^j+\tfrac{1}{16}S^{ij}$} \\
\cmidrule(lr){2-3}\cmidrule(lr){4-5}
\cmidrule(lr){6-7}
    & $m_{SP}^{\rm pot,p}$ & $m_{SP}^{\rm pot,n}$
    & $m_{SP}^{\rm pot,p}$ & $m_{SP}^{\rm pot,n}$ & $m_{SP}^{\rm pot,p}$ & $m_{SP}^{\rm pot,n}$\\
\midrule
 GCN5082 &  -1729 & 1537 & 326.5 & -75.45 &  -1708 & 1532 \\
\midrule
 QX  & -1714 & 1537 & 331.8 & -83.28 & -1694 & 1532 \\
\midrule
Sn100pn & -1757 & 1537 & 324.5 & -22.12 & -1736 & 1535 \\
\hline\hline
\end{tabular}
}
\caption{Results for $m^{\rm pot,p}_{SP}$ and $m^{\rm pot,p}_{SP}$ in $^{138}$Ba in units of $( \mathrm{fm}/e)$. The Gamow-Teller and tensor components of the NME are given separately.}
\label{tab:calc_Msp_potential}
\end{table}

Table~\ref{tab:calc_Msp_potential} presents the results of the shell-model calculations for $^{138}$Ba for the NMEs corresponding to each component of the operator in Eq.~\eqref{eq:Osp_pot_dp_dn}, that is, $m_{SP}^{\rm pot,p}$ and $m_{SP}^{\rm pot,n}$ defined as
\begin{equation}
M_{SP}^{\rm pot}=m_{SP}^{\rm pot,p}\,d_p+m_{SP}^{\rm pot,n}\,d_n\,.    
\end{equation}
The results in Table~\ref{tab:calc_Msp_potential} indicate that, in general, $m^{\rm pot}_{SP}$ NMEs are significantly larger than ultrasoft NMEs. This difference arises from the coherent contribution of all nucleons, in contrast to the non-coherent ultrasoft NME, which is dominated by a few components where the core does not contribute. Coherence also leads to very similar NMEs across the three shell-model Hamiltonians used, indicating that nuclear structure details are not very relevant for the potential NME. Indeed, for $m^{\rm pot,n}_{SP}$ the results are almost identical because in our calculations the 82 neutrons in $^{138}$Ba form a closed shell. 

In addition, Table~\ref{tab:calc_Msp_potential} distinguishes the results for the Gamow-Teller and tensor spin structures. For both neutron and proton parts, the contribution of the tensor is very small. This suggests that the potential NME is mostly driven by pairs of nucleons coupled to spin zero, just as dictated by the Gamow-Teller operator.

\begin{table}[t]
\scalebox{1.0}{
\centering
\begin{tabular}{c cccc }
\hline\hline
\addlinespace[0.8ex]
    & \multicolumn{2}{c}{$V(r)=1$} & \multicolumn{2}{c}{$V(r)=r$} \\
\cmidrule(lr){2-3}\cmidrule(lr){4-5}
    & $m_{SP}^{\rm pot,p}$ & $m_{SP}^{\rm pot,n}$ & $m_{SP}^{\rm pot,p}$ & $m_{SP}^{\rm pot,n}$ \\
\midrule
$^{16}$O   & -66.96  & 45.84  & -244.5  & 167.4 \\
\midrule
$^{20}$Ne  & -82.15  & 56.23   & -303.7 & 207.8 \\
\midrule
$^{36}$S   & -115.3  & 114.6  & -403.7  & 439.8 \\
\midrule
$^{48}$Ca  & -167.4  & 141.1  & -667.8  & 484.6 \\
\midrule
$^{100}$Sn & -367.8 & 250.0  & -1274 & 849.3 \\
\midrule
$^{132}$Sn & -367.2  & 425.9  & -1317 & 1502 \\
\midrule
$^{138}$Ba & -434.5 & 427.3  & -1708 & 1532 \\
\hline\hline
\end{tabular}
}
\caption{Results for $m^{\rm pot,p}_{SP}$ and $m^{\rm pot,n}_{SP}$ in fm/$e$, labeled as $V(r)=r$, for several nuclei across the nuclear chart. In addition to $^{138}$Ba (see main text), for $^{20}$Ne and $^{36}$S we use an $^{16}$O core solved with the USDA \cite{Richter:2008} interaction in the $sd$ shell, and for $^{48}$Ca we take a $^{40}$Ca core solved with the KB3G~\cite{Poves:2001} interaction in the $pf$ shell. $^{16}$O, $^{100}$Sn and $^{132}$Sn are described by a single Slater determinant. We also include results for the NMEs obtained without the radial part of the operator in units of $1/e$, which we label as $V(r)=1$.}
\label{tab:calc_Msp_pot_A}
\end{table}

\begin{figure}[t]
    \centering
    \includegraphics[width=0.45\textwidth]{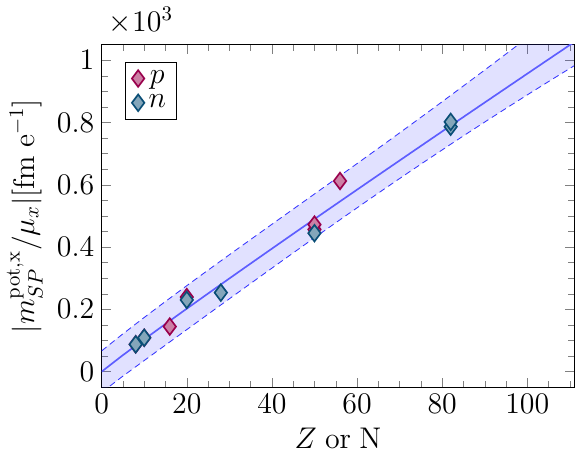} 
    \caption{Absolute value for the proton ($m^{\rm pot,p}_{SP}$) and neutron ($m^{\rm pot,n}_{SP}$) contributions to $M^{\rm pot}_{SP}$, divided by the corresponding nucleon MDM, as a function of the atomic or neutron number. The results cover all nuclei in Table~\ref{tab:calc_Msp_pot_A} and also show the best linear fit (see text) and $95\%$CL prediction bands.}
    \label{fig:Msp_scaling}
\end{figure}
We explore the scaling of the potential NMEs with the number of protons and neutrons by calculating $M^{\rm pot}_{SP}$ for several nuclei in different mass regions. Table~\ref{tab:calc_Msp_pot_A} lists the results, which suggest that the potential NME indeed increases linearly with $Z$ and $N$. Figure~\ref{fig:Msp_scaling} highlights this linear relation. It represents, for all nuclei, $m^{\rm pot,p}_{SP}$ and $m^{\rm pot,n}_{SP}$, normalized by the proton and neutron magnetic moment, as a function of $Z$ ($m^{\rm pot,p}_{SP}/\mu_p$) or $N$ ($m^{\rm pot,n}_{SP}/\mu_n$). A good linear relation common to the proton and neutron parts of all nuclei emerges, best fitted to $m^{\rm pot,X}_{SP}=-(9.74\pm 0.19)\,X\,\mu_X (\mathrm{fm}/e)$, where $X$ stands either for protons ($Z, p$) or neutrons ($N, n$). The uncertainty is purely based on the fit to the performed calculations, but does not contain larger uncertainties from nucleus-dependent and chiral corrections. These are commented on below. The linear relation confirms that the potential NMEs is largely dominated by spin-zero pairs of protons and neutrons, as it includes nuclei where all nucleons form pairs---such as $^{16}$O, $^{36}$S (neutrons) or $^{48}$Ca (protons)---because they fill angular-momentum-closed shells. The contribution of proton-neutron pairs is minor.     

Using the scaling function for the NMEs, we write a master formula one can use to compute the \textit{equivalent} electron EDM $d_e^\text{equiv} \equiv r_\text{mol} \bar C_\text{SP}/A$ \cite{Pospelov:2013sca} for any system
 \begin{equation}
    {d}^{\rm equiv}_e=\frac{\sqrt{2}e^4m_e}{18\pi G_Fm_N} \frac{r_\text{mol}}{A} (-9.74) \left[  \, Z \, \mu_p\, d_p+N \,\mu_n \,d_n \right] \frac{\mathrm{fm}}{e} 
    \label{eq:master formula}
\end{equation}
where $A$ is the mass number of the heaviest nucleus in each system,  and $r_\text{mol}$ is a molecular matrix element. For several molecules of experimental interest it is given by \cite{Haase_2021,Fleig:2018bsf}:
\begin{eqnarray}
r_\text{BaF}&=&\,4.46[18]\cdot 10^{-21} \text{ e cm}\; ,\nonumber\\
    r_\text{ThO}&=&\,1.51[9] \cdot 10^{-20} \text{ e cm}\; ,\nonumber\\
        r_{\text{HfF}^+}&=&9.17[52] \cdot 10^{-21} \text{ e cm}\,.
\end{eqnarray}

Additionally to the full potential operator, indicated by $V(r)=r$, Table~\ref{tab:calc_Msp_pot_A} also presents NMEs for the operator just keeping the spin and isospin degrees of freedom, but without radial dependence. These results, denoted by $V(r)=1$, also indicate a linear dependence with the number of neutrons and protons. In fact, for nucleons in fully angular-momentum-closed shells the NMEs exactly fit to $-3\,X\,\mu_X d_X\; (\mathrm{fm}/e)$, as expected. For all the nuclei in Table~\ref{tab:calc_Msp_pot_A}, the results for $m^{\rm pot,p}_{SP}$ and $m^{\rm pot,n}_{SP}$ share a similar relation with the ones obtained for $V(r)=1$, with a proportionality constant $\sim(3.4-3.9)$~fm. This common factor reveals the lack of additional scaling in the potential NMEs due to the radial part of the operator. 

\begin{figure}[t]
    \centering
    \includegraphics[width=0.45\textwidth]{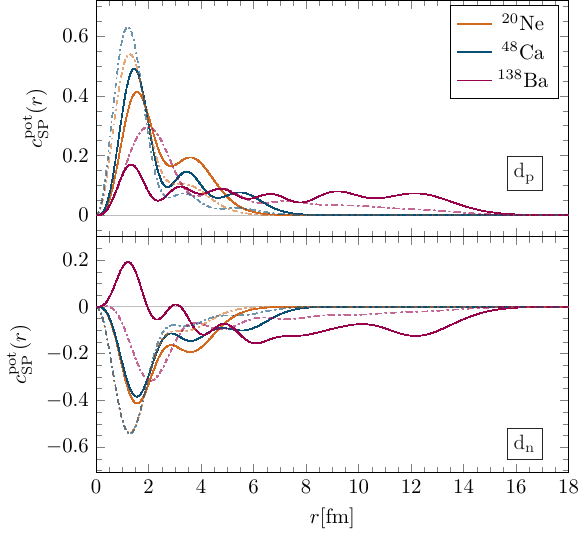} 
    \caption{Normalized $c^{\rm pot}_{SP}(r)$ in units of fm${}^{-1}$ for $^{20}$Ne, $^{48}$Ca, and $^{138}$Ba for the contributions of the proton and neutron EDMs in Eq.~\eqref{eq:Osp_pot_dp_dn}. The dashed lines represent the normalized distributions with no radial dependence.}
    \label{fig:Msp_radial_dist_Vr}
\end{figure}

Figure~\ref{fig:Msp_radial_dist_Vr} further analyzes this aspect, showing the normalized radial distribution $c_{SP}^{\rm pot}(r)$ for $^{20}$Ne, $^{48}$Ca, and $^{138}$Ba, defined by
\begin{equation}
    c_{SP}^{\rm pot}(r)=\frac{\sum_{i\neq j}\mu^{(i)}D^{(j)}r_{ij}N^{(ij)}(r_{ij})\delta(r-r_{ij})}{|M^{\rm pot}_{SP}|}\;,
    \label{eq:Csp_pot_rad}
\end{equation}
which fulfills the relation
\begin{equation}
    1=\int_0^{\infty}c^{\rm pot}_{SP}(r){\rm d}r\;.
    \label{eq:Csp_pot_int}
\end{equation}
The radial distributions in Fig.~\ref{fig:Msp_radial_dist_Vr} show that, regardless of the size of the nucleus, the dominant contribution to the potential NME, shown in solid curves, stems from nucleons relatively close to each other. This property is dictated by the spin part of the NME, indicated by the dashed curves in Fig.~\ref{fig:Msp_radial_dist_Vr}. Even though the full potential operator---including the radial part---gives more relevance to nucleons further apart, pairs of nucleons at short distances still dominate. 
In light of this, we expect that nucleus-dependent effects can correct the estimated slope in Eq.~\eqref{eq:master formula} up to 20$\%$. Higher-order chiral corrections are expected to give a similar uncertainty, leading to a total expected error of about 30$\%$.